\documentclass[twocolumn,showpacs,preprintnumbers,nofootinbib,prd,aps,
superscriptaddress,10pt]{revtex4-1}

\usepackage{amsmath,amssymb}
\usepackage[normalem]{ulem}
\usepackage{textcomp}
\usepackage{hyperref}
\usepackage{bm}
\usepackage{graphicx}
\usepackage{multirow}
\usepackage{subcaption}
\usepackage{array}
\usepackage{psfrag}
\usepackage[usenames,dvipsnames]{xcolor}
\usepackage[utf8]{inputenc}
\usepackage[english, capitalise]{cleveref}
\crefname{chapter}{Chap.}{Chap.}
\crefname{section}{Sec.}{Sec.}
\Crefname{chapter}{Chapter}{Chapters}
\Crefname{section}{Section}{Sections}
\Crefname{eqs}{Eqs.}{Eqs.}		
\Crefformat{eqs}{Eqs.~(#2#1#3)}	

\definecolor{darkgreen}{rgb}{0,0.5,0}
\hypersetup{
	unicode=false,          
	pdftoolbar=true,        
	pdfmenubar=true,        
	pdffitwindow=false,     
	pdfstartview={FitH},    
	pdftitle={NSBH memory},
	pdfauthor={S.~Tiwari, M.~Ebersold, M.~Haney},
	pdfsubject={Gravitational Waves},
	pdfkeywords={gravitational waves, memory effect, general relativity},
	pdfnewwindow=true,      
	colorlinks=true,       
	linkcolor=red,          
	citecolor=cyan,        
	filecolor=magenta,      
	urlcolor=darkgreen,           
	linktocpage=true
}

\newcommand{\Mtot}{M_\mathrm{tot}}

\allowdisplaybreaks[4]

\begin{document}

\title{Observational limits on the rate of radiation-driven binary black hole capture events}

\date{\today}

\author{Michael Ebersold}
\affiliation{Physik-Institut, Universit\"at Z\"urich, Winterthurerstrasse 190, 8057 Z\"urich, Switzerland}

\author{Shubhanshu Tiwari}
\affiliation{Physik-Institut, Universit\"at Z\"urich, Winterthurerstrasse 190, 8057 Z\"urich, Switzerland}

\author{Leigh Smith}
\affiliation{SUPA, School of Physics \& Astronomy, University of Glasgow, Glasgow G12 8QQ, United Kingdom}

\author{Yeong-Bok Bae}
\affiliation{Center for Theoretical Physics of the Universe, Institute for Basic Science (IBS), Daejeon 34126, Korea}

\author{Gungwon Kang}
\affiliation{Department of Physics at Chung-Ang University, Seoul 06974, Korea}

\author{Daniel Williams}
\affiliation{SUPA, School of Physics \& Astronomy, University of Glasgow, Glasgow G12 8QQ, United Kingdom}

\author{Achamveedu Gopakumar}
\affiliation{Tata Institute of Fundamental Research, Mumbai 400005, Maharashtra, India}

\author{Ik Siong Heng}
\affiliation{SUPA, School of Physics \& Astronomy, University of Glasgow, Glasgow G12 8QQ, United Kingdom}

\author{Maria Haney}

\affiliation{Nikhef, Science Park, 1098 XG Amsterdam, Netherlands}

\begin{abstract}
Dense astrophysical environments like globular clusters and galactic nuclei can host hyperbolic encounters of black holes which can lead to gravitational-wave driven capture. There are several astrophysical models which predict a fraction of binary black hole mergers to come from these radiation-driven capture scenarios. In this paper we present the sensitivity of a search toward gravitational-wave driven capture events for O3, the third observing run of LIGO and Virgo.
We use capture waveforms produced by numerical relativity simulations covering four different mass ratios and at least two different values of initial angular momentum per mass ratio. We employed the most generic search for short-duration transients in O3 to evaluate the search sensitivity in this parameter space for a wide range in total mass in terms of visible spacetime volume.
From the visible spacetime volume we determine for the first time the merger rate upper limit of such systems. The most stringent estimate of rate upper limits at 90\% confidence is $0.2~\mathrm{Gpc}^{-3}\,\mathrm{yr}^{-1}$ for an equal mass $200~M_\odot$ binary.
Furthermore, in recent studies the event GW190521 has been suggested to be a capture event.
With this interpretation of GW190521, we find the merger rate of similar events to be $0.47~\mathrm{Gpc}^{-3}\,\mathrm{yr}^{-1}$.

\end{abstract}


\maketitle

\section{Introduction} \label{sec:intro}

Since the first direct detection of gravitational waves (GWs) in 2015~\cite{GW150914-2016}, the ground-based detector network of Advanced LIGO~\cite{LIGO-2014}, Virgo~\cite{Virgo-2014} and KAGRA~\cite{KAGRA-2019} has found almost 100 gravitational wave events~\cite{gwtc-1-2018,gwtc-2-2020,gwtc-3-2021}. The majority are binary black hole (BBH) mergers, but the ground-based detectors have also observed two neutron star coalescences~\cite{GW170817-2017,GW190425-2020}, two neutron star-black hole mergers~\cite{NSBH-2021} and an event with a large mass ratio, where the nature of the secondary object is unclear~\cite{GW190814-2020}. These signals are all consistent with quasicircular compact binary mergers.

The usual searches for stellar-mass compact binary coalescences (CBCs) in quasicircular orbits are conducted using the modeled search algorithms based on matched-filtering~\cite{usman-2015,messick-2016}, this is made possible due to the availability of fast and accurate gravitational waveforms in this parameter space~\cite{cotesta-2018,varma-2019,pratten-2020}.
In addition to these searches, there are dedicated searches conducted for intermediate mass black hole (IMBH) binaries~\cite{o3-imbh-2022} and eccentric binary black holes (eBBH)~\cite{o2-eccentric-2019,ramos-buades-2020}, here the matched-filter and unmodeled methods dedicated toward finding BBH events are both employed due to the short-duration of the signal and waveform model systematics. Indeed the heaviest BBH merger detected till date, GW190521~\cite{GW190521-2020, GW190521-impl-2020}, was found with a dedicated unmodeled search pipeline for BBH events with the highest significance~\cite{cwb_190521}. The component masses of GW190521 were estimated under the assumption of a quasicircular merger to be $m_1 \simeq 85\, M_\odot$ and $m_2 \simeq 66\, M_\odot$ in an orbit with signatures of orbital precession. However, due to the short duration and lack of power at lower frequencies before merger, GW190521 has also allowed for alternative interpretations~\cite{gamba-2021,romero-shaw-2020,nitz-2020,estelles-2021,gayathri-2020,bustillo-2020,bustillo-2021}.

Apart from the CBCs, there are several other types of potential transient GWs signal sources, e.g. supernovae~\cite{szczepanczyk-2021}, bursts of nonlinear memory~\cite{ebersold-2020}, cosmic strings~\cite{cosmicstring-2021}, long lived emission from neutron stars~\cite{thrane-2010}, where the signal model is not well known. Searches for these GW transients with minimal assumptions on the signal morphology, sky-direction and polarization were conducted for both, signals with short~\cite{o3-all-sky-short-2021} and long duration~\cite{o3-all-sky-long-2021}. Here, no significant event was found besides the known CBC events.


The detection of each CBC system improves the insights about how such systems are formed in nature, especially if orbital properties like spin-precession or eccentricity are observed that constitute unique signatures about the formation channel of the binary.
There are several formation channels which predict gravitationally bound BBHs that can merge within a Hubble time, a short review of these formation channels can be found in Ref.~\cite{mapelli-2020}. Two broad classification of formation channels are isolated evolution and dynamical evolution. For isolated evolution the binaries are formed without interaction with other objects or environments and the BBH is formed from the stars. If the binary stars are close they can undergo mass transfer, common envelope, tides and natal kicks to form a tightly bound BBH~\cite{dominik-2012, giacobbo-mapelli-2018}. 
The long evolution time and the isolated nature of this formation channel predicts almost quasicircular merger of BBH~\cite{peters-1963,peters-1964}. It should be noted that there can also be triple isolated systems, where the BBH merger can have eccentricities due to third body interaction~\cite{antonini-2017}. 

The second formation channel is the dynamical evolution scenario. Here dynamical interaction of a single or a binary system with another astrophysical object or the environment can lead to formation of BBHs. There are also several scenarios which predict formation of gravitationally bound BBHs through this channel which are mostly limited to dense astrophysical environments such as globular clusters (GC) and galactic nuclei (GN). The usual timescales from the formation to merger for this formation channel can be as low as few milliseconds i.e. the binary is formed so tightly that it merges without having an inspiral. In this work we focus on this particular formation channel as it is the only scenario where radiation-driven capture, which results in the merger of the BBH within seconds after formation, is possible.


In GN, near the supermassive black hole (SMBH), radiation-driven capture can form very tight binaries by single-single interactions, which typically merge before a third object could alter their orbits~\cite{gondan-2017,gondan-2020,samsing-2020}. Especially for high SMBH masses and large component masses, the binary can form with significant eccentricity even close to merger due to shorter time from formation to merger~\cite{gondan-2017}. It is estimated that between 26\%-50\% of radiation-driven capture events from these single-single interactions will form above 10Hz with high eccentricity, which lies in the LIGO/Virgo/KAGRA detectable band~\cite{gondan-2020}. In GCs, along with single-single interactions~\cite{samsing-2019}, radiation-driven captures can also happen via binary-single~\cite{samsing-2017} and binary-binary~\cite{zevin-2018} interactions. Binary-single capture events are most common in GCs, accounting for roughly 10\% of all BBH mergers formed in these environments~\cite{samsing-2017}. Binary-binary events occur less frequently, however contribute to 25\%-45\% of eccentric mergers that occur during strong black hole (BH) encounters~\cite{zevin-2018}. It is suggested that single-single interactions in GNs produce the highest rate of eccentric capture events~\cite{oleary-2008,rasskazov-2019,gondan-2020}, as encounters are parabolic and form within the frequency band of ground-based detectors. In this work we will interpret our results in the light of the formation channel described in Refs.~\cite{rasskazov-2019,gondan-2020} as this formation channel will have the highest probability of getting an event which merges within seconds after formation. We also note that there could be single-binary encounters in nuclear star clusters, however, it was recently pointed out that they might not be an interesting source for ground-based detectors~\cite{elena-2022}.




In contrast to gravitational waveform models for quasicircular binaries, accurate waveform models for capture scenarios are rather scarce. The most accurate waveforms are computed by numerical relativity (NR), however, these always have the drawback to only cover one particular point in the parameter space. NR waveforms for direct capture events have been described in Refs.~\cite{bae-2017,bae-2020,rueter-2021}. A full model based on the effective-one-body approach for the multipolar merger-ringdown waveform from dynamical capture BH mergers with arbitrary mass ratio and nonprecessing spins is presented in Ref.~\cite{nagar-2020,nagar-1-2021,nagar-2-2021}.

This paper is organized as follows: \cref{sec:waveforms} describes the capture waveforms used for studying the sensitivity of the search. In~\cref{sec:analysis} we present the analysis and in~\cref{sec:astroimpl} we interpret our results in terms of recent astrophysical models that predict radiation-driven capture events. Section~\ref{sec:conclusion} provides ideas for future improvements of the search and concludes the work.

\section{Capture Waveforms} \label{sec:waveforms}

The orbits of two BHs which are initially unbound can be transformed into bound orbits by emitting GWs during close encounters, which is called a gravitational-wave capture. In this study we consider a subset of capture scenarios, the ones that form very tight, such that they merge within seconds after formation. We disregard the scenario where an eccentric binary is formed that slowly radiates away its eccentricity and eventually merges like a quasicircular one.
The waveforms for the capture scenarios we are investigating here are from NR simulations described in Ref.~\cite{bae-2017}. They are created employing the parabolic approximation, meaning that the emitted radiation from a weakly hyperbolic orbit is the same as that from a parabolic orbit with the same distance at the point of closest approach. The orbits for the NR simulations are chosen such that they correspond to the parabolic orbit in classical dynamics, whose initial orbital energy is zero. For large initial angular momenta $L$, the two BHs would show a fly-by orbit, although they are in fact bound, their apocenter is too far to be simulated. With decreasing $L$ the orbits become tighter until the BHs almost directly merge. The radiated energy in these scenarios is the highest around the boundary between fly-by orbits and the direct merging ones. Henceforth, we only consider orbits that are directly merging or have at most one encounter before the merger.

\begin{figure*}[ht]
    \centering
    \includegraphics[width=\textwidth]{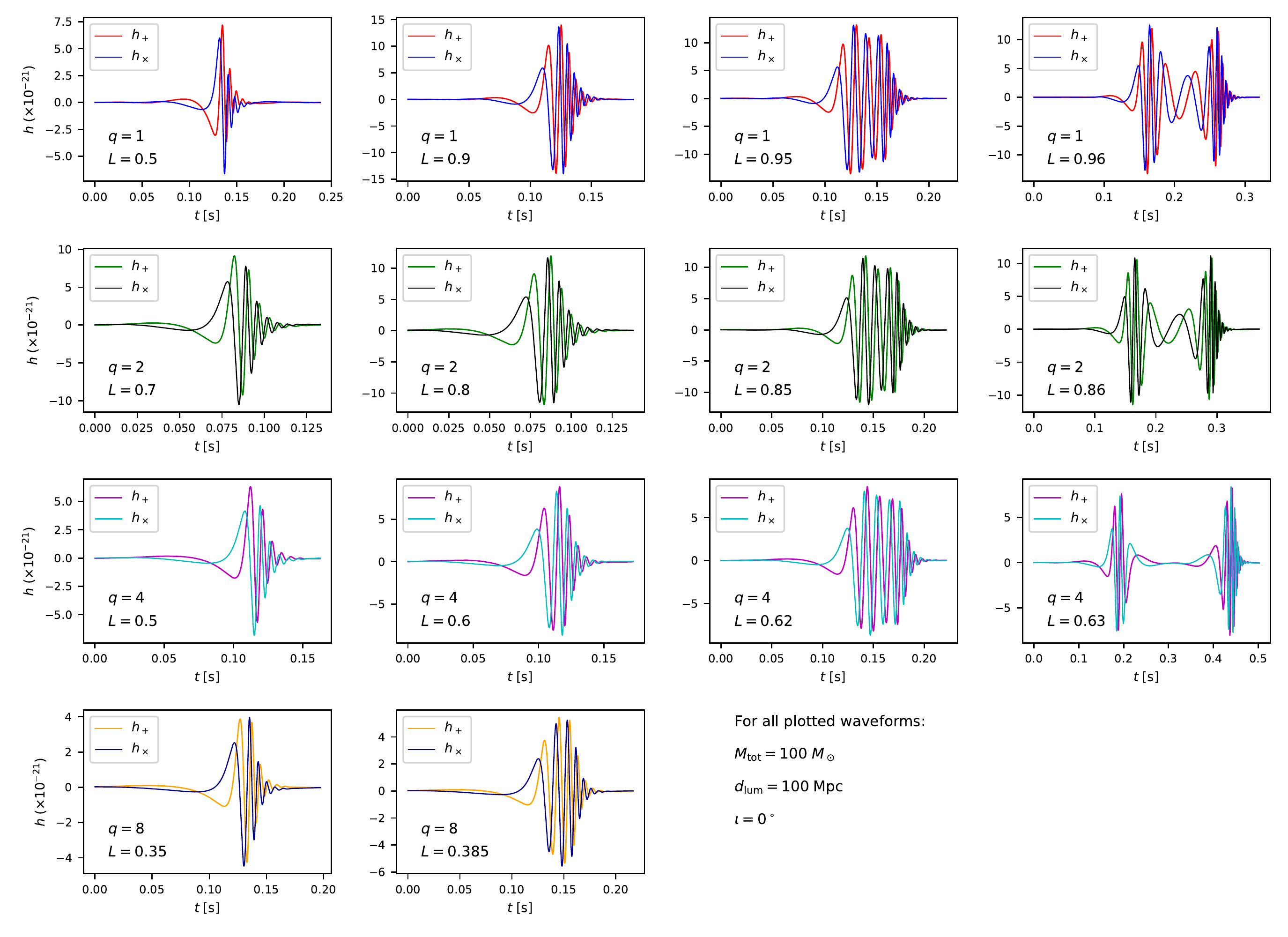}
    \caption{Displayed are the waveform polarizations computed using the dominant 22-mode for all direct capture simulations used in this study. Here they are shown for a face-on binary of total mass $100~M_\odot$ at a distance of 100~Mpc. The top row waveforms are for mass ratio 1, the second row shows mass ratio two, the third row mass ratio 4 and in the bottom row we show the two waveforms for mass ratio 8. The initial angular momentum $L$ and thus the impact parameter is increasing from the left to the right. Note that the initial orbital energies for these simulations are all zero.}
    \label{fig:cap_waveforms}
\end{figure*}

We consider in total 14 different simulations which include four different mass ratios $q = m_1/m_2 = \{1,2,4,8\}$, all having different initial angular momenta $L$. The maximum value of $L$ is always chosen to ensure a capture orbit for each mass ratio. It roughly scales with the reduced mass, implying that the specific initial angular momenta at the boundary are comparable~\cite{bae-2017}. In~\cref{fig:cap_waveforms} we explicitly show the waveforms for a given total mass of $100~M_\odot$. For small initial angular momenta, and therefore small impact parameters, the BHs almost directly merge, the corresponding waveform looks like half a hyperbolic encounter waveform followed by a ringdown. Increasing the initial angular momentum, the BHs start whirling before the merger, which adds additional structure before the ringdown in the waveform. For the highest initial angular momentum we are considering, the BHs after a close encounter zoom somewhat away but merge soon after. In this case, the waveform shows a double blip structure. Note again that we are not considering any potential radiation emitted earlier as the NR waveforms have a limited length.

The NR waveforms provided are decomposed into spin-weighted spherical harmonic modes, with all modes up to $\ell = 4$. The dominant modes are the $\ell = 2$, $m = \pm 2$-modes. The waveforms for the injection sets are computed using only the dominant mode. To check the significance of the contribution of higher modes, we computed the difference in root sum-squared amplitude $h_\mathrm{rss}$ between the waveforms including all modes and only the dominant $22$-mode over all phase and inclination angles. We find that the higher modes affect the waveform the most for high mass ratios and toward edge-on orientation as well as for small values of initial angular momentum. So we find the relative difference in $h_\mathrm{rss}$ averaged over phase and inclination angle in the case of the $q=1$, $L = 0.96$ to be 2.4\%, while for the $q=8$, $L=0.35$ it goes up to 7.3\%. Since the search employed here does not rely on waveform models, the contribution of higher modes to the $h_\mathrm{rss}$ is not significant. However, the effect of higher modes on the phase evolution of the waveform can lead to significant mismatches ($>10$\%). The search algorithm used here can deal well with  higher mode contributions as is shown in~\cite{bustillo-2017,chandra-2020}.  

\begin{figure}[ht]
    \centering
    \includegraphics[width=0.48\textwidth]{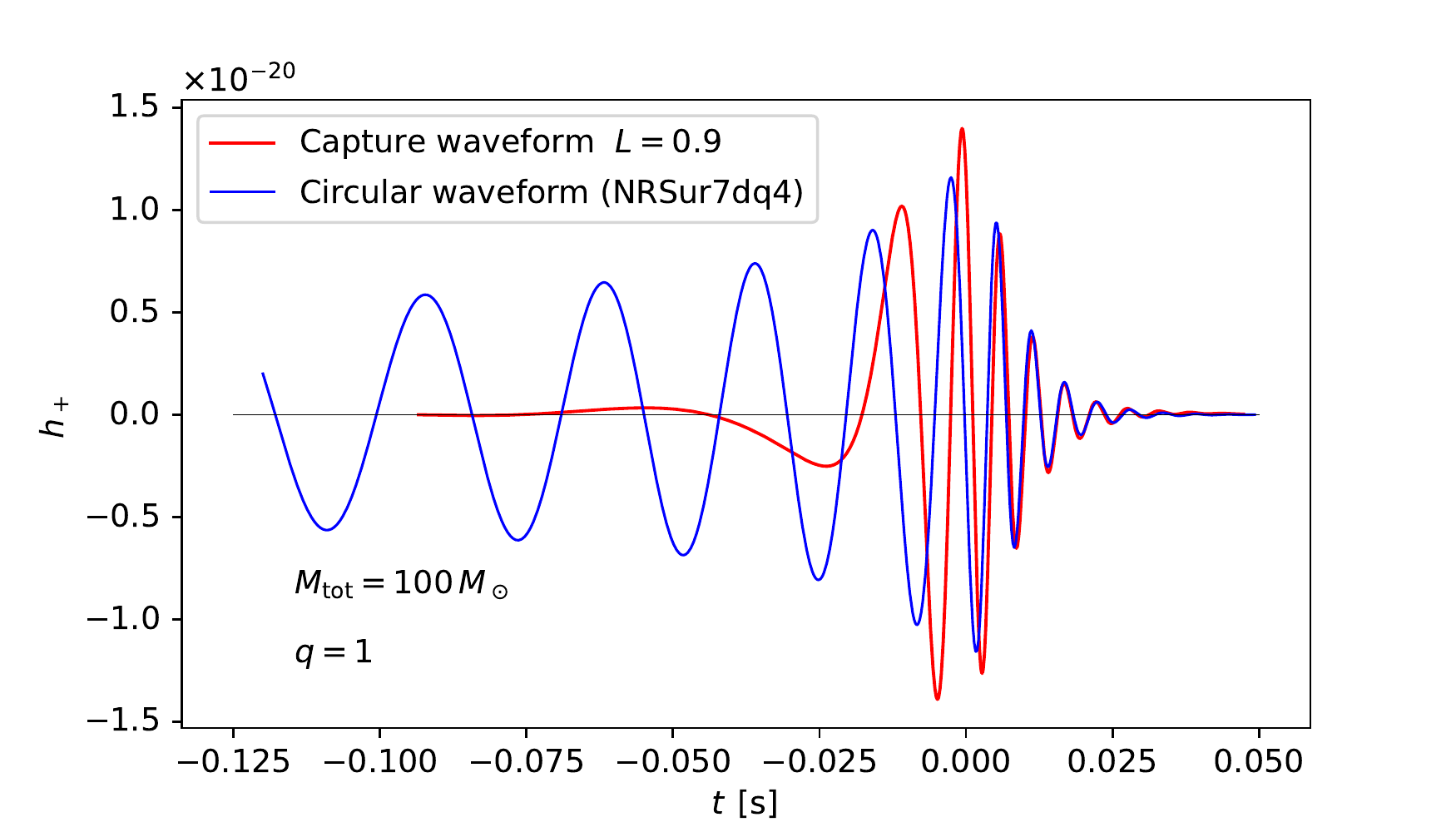}
    \caption{Comparison between a direct capture waveform and the last part of a circular BBH waveform. Both waveforms are nonspinning, the fiducial orientation is face-on at a fixed distance of 100~Mpc.}
    \label{fig:cap_circ}
\end{figure}

To check our waveforms scale correctly with total mass and distance we compared them
against circular templates with the same mass. As the ringdown for nonspinning BHs only depends on the total mass, amplitude and frequency after the merger should be the same. This is indeed the case and can be visually checked in~\cref{fig:cap_circ}, where we display a direct capture waveform together with a circular one generated with the NRSur7dq4 BBH waveform approximant~\cite{varma-2019}. The coalescence phase is chosen such that the overlap is maximized. We can also recognize that the maximal amplitude during the merger is slightly higher in the capture case.

\section{Analysis} \label{sec:analysis}

\subsection{Data} 
The third observing run (O3) of the LIGO-Virgo detectors started on 1 April, 2019 and was concluded on 27 March, 2020. The data taking was paused for about a month from 1 October, 2019 till 1 November, 2019. After the duty cycle consideration, removing periods with poor data quality and requiring each segment of data to be at least 200 seconds we have a total coincident analyzed time of 104.9 days for the first part before the break and 93.4 days for the second part for the Hanford and Livingston detectors. Making the total observation time during O3 in our analysis to be 198.3 days. In this work we have used the publicly available data~\cite{gwosc-2021} from the third observing run of the LIGO-Virgo detectors.

\subsection{Search algorithm}
In this work we have employed a search algorithm called coherent WaveBurst (cWB)~\cite{klimenko-2016,drago-2021}. cWB is a versatile algorithm to search and reconstruct transient GW signals. cWB is an all-sky morphology-independent algorithm i.e. it does not rely on the waveform models or the sky direction of the source. Instead cWB is based on noting the coherent excess power in the network of ground-based detectors as a function of sky direction. cWB has been routinely used in a variety of GW transient searches like IMBH searches~\cite{o3-imbh-2022}, eBBH searches~\cite{o2-eccentric-2019,ramos-buades-2020} and generic searches for transients with short~\cite{o3-all-sky-short-2021} and long duration~\cite{o3-all-sky-long-2021}. The instance of cWB used here is exactly the same which was used for obtaining the results for the short-duration transients with generic morphology in all-sky directions for the third observing run of LIGO-Virgo detectors~\cite{o3-all-sky-short-2021}. We have also used the same techniques for the glitch mitigation, i.e. cuts to remove excess transient noise and classification of search into three bins~\cite{o3-all-sky-short-2021}. We have only considered here the low-frequency analysis between 32 and 1024 Hz as the investigated capture events are expected in this frequency range. We use only the Hanford-Livingston network which was found to be the most sensitive network pair in the analysis.  

The search background is generated from the 198.3 days of coincident data from the two LIGO detectors by time sliding LIGO Hanford with respect to LIGO Livingston by much more than the time of flight between the detectors. The search background corresponds exactly to the one generated in~\cite{o3-all-sky-short-2021}. The significance of each trigger is then computed by noting the detection statistics with respect to the background distribution and is quoted in terms of false alarm rate (FAR) in Hz. In this work we quote the significance as a reciprocal of FAR: inverse false alarm rate (iFAR) in years. The threshold on iFAR for detection of an event is chosen to be $>$ 100 years. The search described above finds a number of significant events which were all identified as the known BBH systems noted in GWTC-2~\cite{gwtc-2-2020,gwtc-2.1-2021} and GWTC-3~\cite{gwtc-3-2021}. After excising the known CBC events from the search, the loudest event was found to have an iFAR of 0.5 years which is well within the expected background rate.

The analysis in Ref.~\cite{o3-all-sky-short-2021} distributed all the triggers into 3 bins. The binning choice was made in order to limit the effect of the very short duration transient glitches which were found to be dominating the background. These glitches were of particular temporal feature and the binning choices were made accordingly i.e. the bins definition was based on the duration of the signal and number of cycles. The bin with highest glitch rate was the one which contained events with a single cycle, the bin with second highest glitch rate contained all the signals with quality factor below 3 and the final bin which had low glitch rate contained all the signals above quality factor 3. In the signal space which we consider here, this binning affects the low angular momentum signals particularly as they are very short and a non negligible fraction of them go into the bins with high glitch rate. The signals with high angular momentum are longer in duration and hence are detected more efficiently. This choice of binning leads to a trials factor of 3 over all the detected events.

The search presented here does find the event GW190521~\cite{GW190521-2020} with an iFAR of 65 years which is significant. We later discuss this event in more details in Sec.~\ref{sec:astroimpl}.


\subsection{Search sensitivity}
Our goal is to estimate the sensitivity of the O3 all-sky search for short-duration transients described in the previous section to gravitational-wave driven captures. We inject waveforms into the data in a broad parameter space. Given the absence of well motivated population models and the limited number of available NR waveforms of direct captures we sample the parameter space in a discrete manner. 

The injection set consists of the 14 waveforms described in~\cref{sec:waveforms}. These are all nonspinning BBH waveforms from NR simulations with different mass ratio and initial angular momentum. For each of the waveforms we consider four different total source masses, namely $M_{\mathrm{tot}} = \{20, 50, 100, 200\}\, M_\odot$ yielding a total of 56 injection sets. The injections are placed uniformly distributed in sky location and inclination angle. The injections are uniformly distributed in comoving volume up to a redshift $z_{\mathrm{max}}^i$ which is chosen differently for each injection set such that no signals from that maximal distance are recovered but also low enough to avoid making injections that are well outside any possible detection range. Each signal is individually redshifted according to the cosmological parameters given in Ref.~\cite{Planck-2018}. Its mass in the detector frame therefore is a factor of $(1+z)$ larger than in the source frame, which shifts it toward lower frequencies.  On the other hand, the luminosity distance is $(1+z)$ times larger than the co-moving distance to the source. The injections are performed over the entirety of O3 data with each injection set having at least 200,000 injections to obtain large statistics in order to have less than 1\% error in detection efficiency estimates as a function of source parameters and redshift. 

The averaged spacetime volume to which our search is sensitive is given by~\cite{tiwari-2017,LIGO-2016-1,LIGO-2016-2}:
\begin{align}
    \langle V T \rangle = \int dz \, d\theta \frac{dV_c}{dz} \frac{1}{1+z} p_\mathrm{pop}(\theta) f(z,\theta) \, T\,.
\end{align}
In this equation $dV_c/dz$ is the differential comoving volume, $p_\mathrm{pop}$ is the distribution of binary parameters $\theta$, $f(z,\theta)$ is the probability of recovering a signal with parameters $\theta$ at redshift $z$ and $T$ is the length of the observation in the detector frame. 

The corresponding sensitive distance reach is computed from the average sensitive spacetime volume as
\begin{align}
    D_\mathrm{\langle V T \rangle} = \left( \frac{3 \langle V T \rangle}{4 \pi T_s} \right)^{1/3}\,,
\end{align}
where $T_s$ is the length of detector data analyzed by the search.
\begin{figure*}[ht]
    \centering
    \begin{subfigure}[b]{0.497\textwidth}
        \centering
        \includegraphics[width=\textwidth]{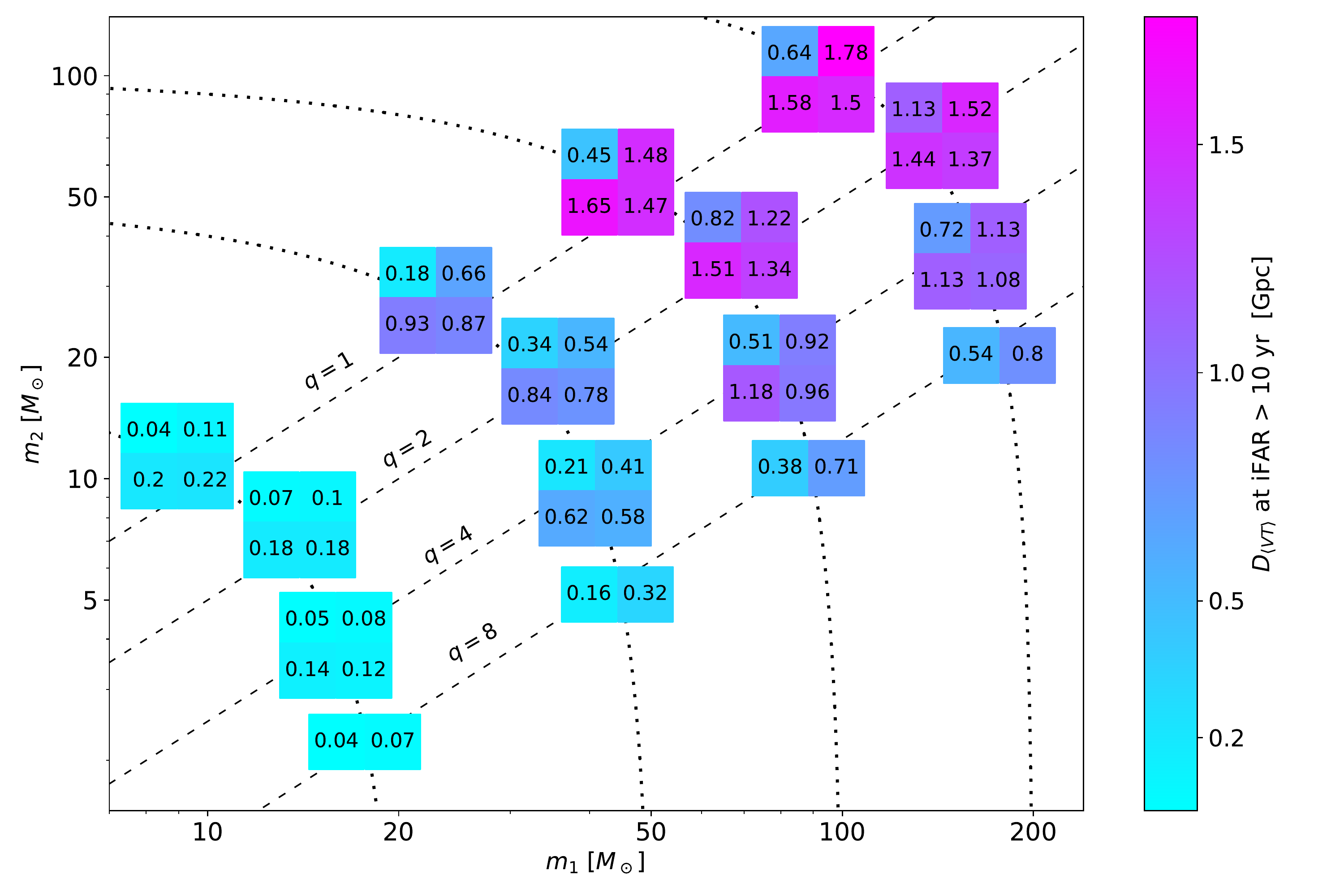}
        \caption{iFAR $>$ 10 yr}
    \end{subfigure}
    \hfill
    \begin{subfigure}[b]{0.497\textwidth}
        \centering
        \includegraphics[width=\textwidth]{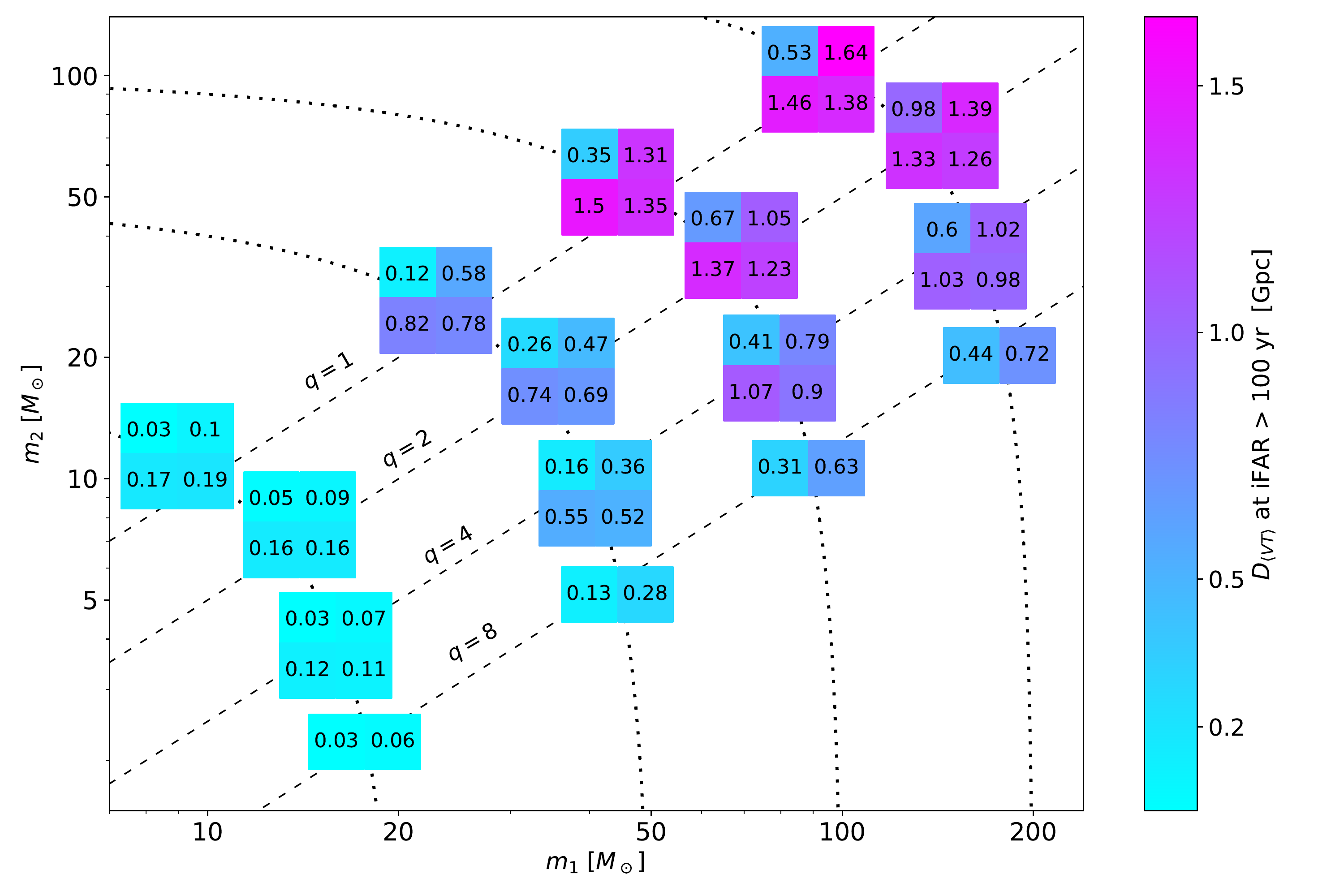}
        \caption{iFAR $>$ 100 yr}
    \end{subfigure}
    \caption{Sensitive distance reach at iFAR $> 10$ yr and $> 100$ yr displayed in the source frame $m_1 - m_2$ plane for all analyzed direct capture waveforms. Each box with 4 (2 for mass ratio 8) values has the same total mass and mass ratio. The results for the different values of initial angular momentum $L$ are arranged as follows: at the top left is the lowest, and it increases toward the top right, bottom left and is the highest at the bottom right.}
    \label{fig:reach}
\end{figure*}
In~\cref{fig:reach} we show the sensitive distance reach for all analyzed direct capture waveforms at two different iFAR thresholds of 10 and 100 years. Even with the most generic search for GW transients, which we have employed here, the surveyed distance can go up to $\approx 1.7$ Gpc for high mass systems in O3 which is not far from the one for the usual quasicircular orbit BBH. This is expected as for high mass systems most of the signal strength is given by the merger-ringdown part for the current detectors, which is similar in both cases as shown in Fig.~\ref{fig:cap_circ}. As we go down in total mass the surveyed distance drops more rapidly as compared to the circular systems mainly due to the lack of inspiral part of the signal in the capture scenario. Interestingly, although the energy emission in GWs increases with higher initial angular momenta $L$, with the maximum being at the border to the fly-by case~\cite{bae-2017}, the sensitive distance reach is in general not the largest for the capture waveform with the highest $L$. This is because also the frequency content of the emitted radiation shifts towards the lower end where the detectors are less sensitive.

\section{Astrophysical implications}
\label{sec:astroimpl}

As discussed in~\cref{sec:intro}, the formation channel which can form the BBH merger events seconds after their formation, is the dynamical evolution scenario. This scenario for BBH formation predicts a fraction of direct capture events with single-single interaction. In this section we will interpret the results obtained in the previous~\cref{sec:analysis} in terms of rates and we will discuss it further following the formation channel considered in Refs.~\cite{rasskazov-2019,gondan-2020}.  

We briefly introduce how the astrophysical rates can be estimated from the sensitive spacetime volume $\langle V T \rangle$. Given $\langle V T \rangle$ of an injection set $i$ at some iFAR threshold $x$, which we refer to as $\langle V T \rangle_{i}^{\mathrm{iFAR} > x}$, the number of expected detections at the same iFAR threshold $N^{\mathrm{iFAR} > x}$ given the astrophysical rate $R$ is given as 
\begin{equation}
    \langle N \rangle^{\mathrm{iFAR} > x} = R \times \langle V T \rangle_{i}^{\mathrm{iFAR} > x}\,.
\end{equation}
We can assume the population of astrophysical source produce a Poisson distribution with rate $R$. In the absence of any detection the upper limit on the rate at 90\% confidence interval is given as~\cite{brady-2004}
\begin{equation}
    R_{90} = \frac{2.303}{ \langle V T \rangle_{i}^{\mathrm{iFAR}>x}} \,.
\end{equation}
\begin{figure}[ht]
    \centering
    \includegraphics[width=0.482\textwidth]{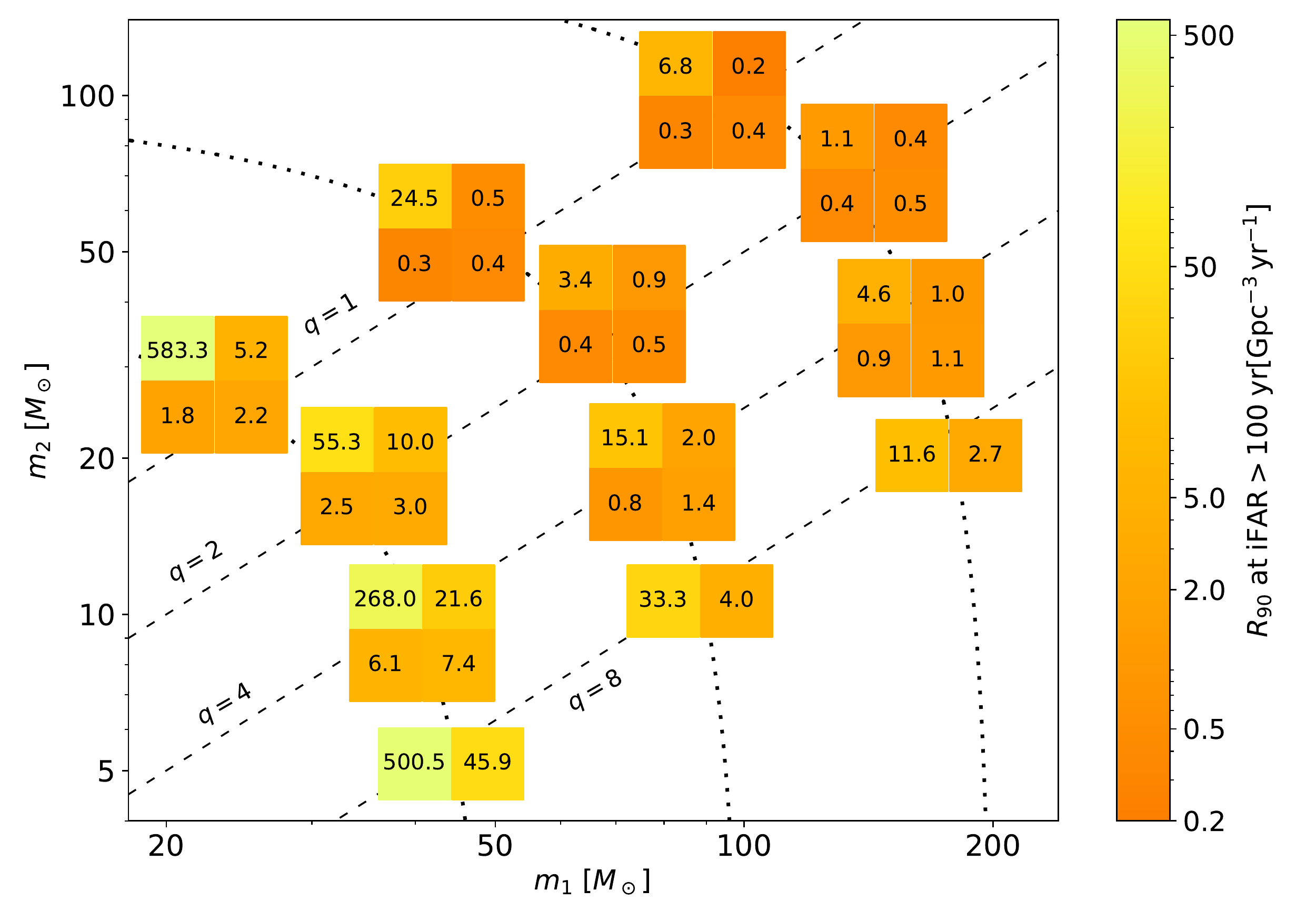}
    \caption{Rate upper limit at iFAR threshold of 100 years for our injections sets with total mass $\{50,\, 100,\, 200\}~M_\odot$. The waveforms in each box are arranged in the same way as described in~\cref{fig:reach}. }
    \label{fig:rates}
\end{figure}

We provide this rate upper limit for our injection sets at an iFAR threshold of 100 years in~\cref{fig:rates}. Here we omit the values for $20~M_\odot$ as they are not competitive. Nonetheless, they can be found in~\cref{tab:tabrange}. The best rate upper limit we get is for an equal mass, $200~M_\odot$ binary, with angular momentum $L=0.9$ at 0.23 $\mathrm{Gpc}^{-3}\mathrm{yr}^{-1}$. For the same total mass this rate upper limit is over an order of magnitude worse than the one presented in Ref.~\cite{o3-imbh-2022}. This is expected because \textit{first} the search presented here is the search for the most generic short-duration transients, \textit{secondly} the rate upper limit obtained is for a system with same mass but long inspiral due to aligned spins.
In light of the astrophysical model that can produce these events, described in Ref.~\cite{rasskazov-2019}, the rate of events are expected to be $0.002-0.04$ $\mathrm{Gpc}^{-3}\mathrm{yr}^{-1}$ which is still not achieved by the current analysis in third observing run by almost an order of magnitude. In the ``gas-driven" capture scenario presented in Ref.~\cite{tagawa-2019}, where the event rate can be $0.02-60$ $\mathrm{Gpc}^{-3}\mathrm{yr}^{-1}$, a fraction of events could actually be detected from this formation channel. This formation channel also predicts high masses with high eccentricity mergers which can be a possibility for GW190521. 

\begin{table}[ht]
    \centering	\setlength\extrarowheight{4pt}
    \begin{tabular}{| c | c | c | c | c | c | c|}
        \hline 
		\: $\Mtot$ \: & \: $q$ \: & \: $L$ \:\: & \: $z_\mathrm{max}$ \: & \: $\langle VT \rangle$ \: & \: $D_{\langle VT \rangle}$ \: & \: $R_{90}$ \:  \\
		{\small $[M_\odot]$} & & & & {\small $[\mathrm{Gpc}^3 \, \mathrm{yr}]$} & {\small$[\mathrm{Gpc}]$} & {\small $[\frac{1}{\mathrm{Gpc}^{3}\, \mathrm{yr}}]$} \\[1ex]
		\hline
		\multirow{8}{*}{100} & \multirow{4}{*}{1} & 0.5 & 0.32 & 0.11 & 0.36 & 36.7 \\ 
			& & 0.9 & 1.02 & 5.56 & 1.35 & 0.70 \\ 
			& & 0.95 & 1.12 & 8.36 & 1.54 & 0.47 \\ 
			& & 0.96 & 1.07 & 6.03 & 1.38 & 0.65 \\ 
			\cline{2-7}
			 & \multirow{4}{*}{2} & 0.7 & 0.8 & 0.76 & 0.69 & 5.16 \\ 
			& & 0.8 & 0.93 & 2.91 & 1.09 & 1.34 \\ 
			& & 0.85 & 1.03 & 6.33 & 1.41 & 0.62 \\ 
			& & 0.86 & 0.93 & 4.59 & 1.26 & 0.85 \\ 
			\hline
    \end{tabular}
    \caption{For the injections sets close to the estimated mass and mass ratio of GW190521 the table shows the surveyed spacetime volume, sensitive distance reach and rate at an iFAR threshold of 65 years.}
    \label{tab:GW190521}
\end{table}

 We have two hypothesis due to lack of unambiguous understanding of the nature of the event GW190521. The two hypotheses are 
\begin{itemize}
    \item GW190521 is indeed a GW driven capture event and hence in this case we evaluate the rate of this source class. 
    \item GW190521 is not a GW driven capture event and in this case the evaluated rate upper limits at iFAR threshold of 100 years can be found in~\cref{fig:rates}.  
\end{itemize}
For the case of GW190521 as a capture event, a detailed analysis was conducted in Ref.~\cite{gamba-2021}, where the estimated total mass of the system was found to be $130^{+75}_{-43}\;M_{\odot}$ and the mass ratio has support at both, $q \approx 1$ and $q \approx 2$. We choose the injection sets with total mass $100~M_{\odot}$ and mass ratios $\{1,\,2\}$, which is a conservative choice since for the injection set with total mass $200~M_\odot$ the distance reach increases, although only slightly. The sensitive volume for the suitable injection sets at $\mathrm{iFAR} > 65~\mathrm{yr}$ are stated in~\cref{tab:GW190521} and are about a few $\mathrm{Gpc}^3\, \mathrm{yr}$. Assuming this to be one event, the rate can be computed as in Ref.~\cite{o3-imbh-2022}
\begin{align}
    R_{90} = \frac{3.9}{ \langle V T \rangle_{i}^{\mathrm{iFAR} > 65 \mathrm{yr}}}\,,
\end{align}
which yields $0.47~\mathrm{Gpc^{-3}}\,\mathrm{yr^{-1}}$ for mass ratio 1 and $0.62~\mathrm{Gpc^{-3}}\,\mathrm{yr^{-1}}$ for mass ratio 2, in the case of the waveform which can be seen the farthest. These numbers are in agreement with the rate estimates given in Ref.~\cite{tagawa-2019}, however due to lack of unambiguous understanding of the nature of GW190521 a definitive statement cannot be made. Also, the detection of a single event is not enough to validate a particular formation channel. In the future observing runs with improved analysis and upgraded detectors, there can be a population of such events which can then provide a conclusive understanding regarding the origin of such binaries.

\section{Conclusion} \label{sec:conclusion}

In the present work we have for the first time estimated the average visible volume of GW driven capture events using NR simulations. We have employed the most generic transient search for GW signals in the low frequency part where the source is expected. The visible spacetime volume indicates that these events can be probed as far away as the usual circular binaries especially for the high masses. We have computed the rates upper limits for a wide parameter space in component masses and angular momentum and have computed the rate of GW190521-like events assuming this event was actually a GW driven capture. 

It should be noted that in this work we have employed the most generic search for GW transients from all-sky directions without any further tuning of the algorithm specific to the source. There are several means based on Machine Learning algorithms like XGboost~\cite{mishra-2021} or GMM~\cite{lopez-2021} which can be employed to have a dedicated search for a better distinction of signals from this source and detector glitches. Further improvements to the cWB algorithm is also envisaged which can lead to a better collection of signal energy to detect these events more efficiently~\cite{klimenko-2022}. Furthermore, since the NR waveforms used here are limited to nonspinning components, the parameter space can be further expanded by the addition of spins to the NR waveforms.

\section*{Acknowledgments}

We thank Antoni Ramos-Buades and Marco Drago for their crucial support in setting up the injection infrastructure. We also thank Sergey Klimenko for pointing us to the inconsistencies in the  amplitude scaling, which lead to further studies and corrections to the injected waveform.

We would like to thank Juan Calderon Bustillo for his detailed comments on an early version of this manuscript.
We would like to thank Hyung Mok Lee for the kind invitation to KASI, Daejon, South Korea where this project was first conceived. We would also like to thank Giovanni Prodi, Edoardo Milotti and Francesco Salemi for discussion and encouragement throughout this project. A.G. would like to acknowledge the Pauli Center for Theoretical Physics and Philippe Jetzer for their kind invitation to the University of Zurich, Switzerland. 
        
M.E. acknowledges support from Swiss National Science Foundation (SNSF) grant No. 200020-182047. S.T. is supported by Swiss National Science Foundation (SNSF) Ambizione Grant No. : PZ00P2\_202204. L.S. is supported by the Science and Technology Facilities Council [ST/V506692/1 2446745]. Y.B.B. was supported by IBS under Project Code No. IBS-R018-D1 and by the National Research Foundation of Korea(NRF) grant funded by the Korea government(MSIT) (No. NRF-2021R1F1A1051269).
G.K. and Y.B.B. were supported by the National Supercomputing Center with supercomputing resources including technical support (KSC-2019-CRE-0256 \& KSC-2020-CRE-0352). G.K. is supported in part by National Research Foundation of Korea (NRF2021R1A2C201247312 ). 
D.W. and I.S.H. were supported by Science and Technology Facilities Council (STFC) grants ST/V001736/1 and ST/V005634/1, and the European Cooperation in Science and Technology (COST) action CA17137.
M.H. acknowledges support from the SNSF grant No. IZCOZ0\_177057 and the University of Zurich's Forschungskredit.

This research has made use of data obtained from the Gravitational Wave Open Science Center (\href{https://www.gw-openscience.org}{https://www.gw-openscience.org}), a service of LIGO Laboratory, the LIGO Scientific Collaboration and the Virgo Collaboration. The authors are grateful for computational resources provided by the LIGO Lab (CIT) and supported by National Science Foundation Grants PHY-0757058 and PHY-0823459. 

This material is based upon work supported by NSF’s LIGO Laboratory which is a major facility fully funded by the National Science Foundation.The authors gratefully acknowledge the Italian Istituto Nazionale di Fisica Nucleare (INFN), the French Centre National de la Recherche Scientifique (CNRS) and the Netherlands Organization for Scientific Research (NWO), for the construction and operation of the Virgo detector and the creation and support of the EGO consortium.

\appendix

\section{Table with results}

Here we present in~\cref{tab:tabrange} the injection parameters and results for all injection sets analyzed in this study.

\begin{table*}[ht]
	\setlength\extrarowheight{4pt}
		\begin{tabular}{| c | c | c | c | c | c | c|} 
			\hline 
			\: $\Mtot$ \: & \: $q$ \: & \: $L$ \:\: & \: $z_\mathrm{max}$ \: & \: $\langle VT \rangle$ \: & \: $D_{\langle VT \rangle }$ \: & \: $R_{90}$ \:  \\
			{\small $[M_\odot]$} & & & & {\small $[\mathrm{Gpc}^3 \, \mathrm{yr}]$} & {\small$[\mathrm{Gpc}]$} & {\small $[\frac{1}{\mathrm{Gpc}^{3}\, \mathrm{yr}}]$} \\[1ex]
			\hline
			\multirow{14}{*}{20} & \multirow{4}{*}{1} & 0.5 & 0.03 & 0.00006 & 0.030 & 37641 \\ 
			& & 0.9 & 0.08 & 0.002 & 0.10 & 1002 \\ 
			& & 0.95 & 0.15 & 0.012 & 0.17 & 190.4 \\ 
			& & 0.96 & 0.15 & 0.016 & 0.19 & 145.2 \\ 
			\cline{2-7}
			 & \multirow{4}{*}{2} & 0.7 & 0.04 & 0.0003 & 0.050 & 8074 \\ 
			& & 0.8 & 0.06 & 0.0014 & 0.085 & 1625 \\ 
			& & 0.85 & 0.12 & 0.0093 & 0.16 & 248.6 \\ 
			& & 0.86 & 0.12 & 0.0095 & 0.16 & 241.3 \\ 
			\cline{2-7}
			 & \multirow{4}{*}{4} & 0.5 & 0.03 & 0.00008 & 0.032 & 28715 \\ 
			& & 0.6 & 0.05 & 0.0008 & 0.071 & 2779 \\ 
			& & 0.62 & 0.09 & 0.0040 & 0.12 & 581.0 \\ 
			& & 0.63 & 0.08 & 0.0033 & 0.11 & 690.6 \\ 
			\cline{2-7}
			 & \multirow{2}{*}{8} & 0.35 & 0.025 & 0.00005 & 0.028 & 48546 \\ 
			& & 0.385 & 0.04 & 0.0005 & 0.059 & 4811 \\ 
			\hline
			\multirow{14}{*}{100} & \multirow{4}{*}{1} & 0.5 & 0.32 & 0.094 & 0.35 & 24.5 \\ 
			& & 0.9 & 1.02 & 5.08 & 1.31 & 0.45 \\ 
			& & 0.95 & 1.12 & 7.75 & 1.50 & 0.30 \\ 
			& & 0.96 & 1.07 & 5.58 & 1.35 & 0.41 \\ 
			\cline{2-7}
			 & \multirow{4}{*}{2} & 0.7 & 0.8 & 0.67 & 0.67 & 3.43 \\ 
			& & 0.8 & 0.93 & 2.65 & 1.05 & 0.87 \\ 
			& & 0.85 & 1.03 & 5.85 & 1.37 & 0.39 \\ 
			& & 0.86 & 0.93 & 4.25 & 1.23 & 0.54 \\ 
			\cline{2-7}
			 & \multirow{4}{*}{4} & 0.5 & 0.32 & 0.15 & 0.41 & 15.1 \\ 
			& & 0.6 & 0.62 & 1.14 & 0.79 & 2.02 \\ 
			& & 0.62 & 0.77 & 2.77 & 1.07 & 0.83 \\ 
			& & 0.63 & 0.66 & 1.64 & 0.90 & 1.40 \\ 
			\cline{2-7}
			 & \multirow{2}{*}{8} & 0.35 & 0.28 & 0.069 & 0.31 & 33.3 \\ 
			& & 0.385 & 0.42 & 0.57 & 0.63 & 4.03 \\ 
			\hline
		\end{tabular}
        \qquad
		\setlength\extrarowheight{4pt}
		\begin{tabular}{| c | c | c | c | c | c | c|}
			\hline 
			\: $\Mtot$ \: & \: $q$ \: & \: $L$ \:\: & \: $z_\mathrm{max}$ \: & \: $\langle VT \rangle$ \: & \: $D_{\langle VT \rangle }$ \: & \: $R_{90}$ \:  \\
			{\small $[M_\odot]$} & & & & {\small $[\mathrm{Gpc}^3 \, \mathrm{yr}]$} & {\small$[\mathrm{Gpc}]$} & {\small $[\frac{1}{\mathrm{Gpc}^{3}\, \mathrm{yr}}]$} \\[1ex]
			\hline
			\multirow{14}{*}{50} & \multirow{4}{*}{1} & 0.5 & 0.15 & 0.004 & 0.12 & 583 \\ 
			& & 0.9 & 0.45 & 0.44 & 0.58 & 5.22 \\ 
			& & 0.95 & 0.7 & 1.27 & 0.82 & 1.82 \\ 
			& & 0.96 & 0.65 & 1.07 & 0.78 & 2.15 \\ 
			\cline{2-7}
			 & \multirow{4}{*}{2} & 0.7 & 0.3 & 0.042 & 0.26 & 55.3 \\ 
			& & 0.8 & 0.4 & 0.23 & 0.47 & 9.99 \\ 
			& & 0.85 & 0.56 & 0.93 & 0.74 & 2.47 \\ 
			& & 0.86 & 0.56 & 0.76 & 0.69 & 3.05 \\ 
			\cline{2-7}
			 & \multirow{4}{*}{4} & 0.5 & 0.15 & 0.009 & 0.16 & 268 \\ 
			& & 0.6 & 0.25 & 0.11 & 0.36 & 21.6 \\ 
			& & 0.62 & 0.4 & 0.38 & 0.55 & 6.08 \\ 
			& & 0.63 & 0.4 & 0.31 & 0.52 & 7.41 \\ 
			\cline{2-7}
			 & \multirow{2}{*}{8} & 0.35 & 0.12 & 0.005 & 0.13 & 500 \\ 
			& & 0.385 & 0.2 & 0.050 & 0.28 & 45.9 \\ 
			\hline 
			\multirow{14}{*}{200} & \multirow{4}{*}{1} & 0.5 & 0.42 & 0.34 & 0.53 & 6.79 \\ 
			& & 0.9 & 1.2 & 10.1 & 1.64 & 0.23 \\ 
			& & 0.95 & 1.16 & 7.12 & 1.46 & 0.32 \\ 
			& & 0.96 & 1.16 & 6.01 & 1.38 & 0.38 \\ 
			\cline{2-7}
			 & \multirow{4}{*}{2} & 0.7 & 0.9 & 2.13 & 0.98 & 1.08 \\ 
			& & 0.8 & 1.0 & 6.15 & 1.39 & 0.37 \\ 
			& & 0.85 & 0.92 & 5.30 & 1.33 & 0.43 \\ 
			& & 0.86 & 0.92 & 4.50 & 1.26 & 0.51 \\ 
			\cline{2-7}
			 & \multirow{4}{*}{4} & 0.5 & 0.56 & 0.50 & 0.60 & 4.60 \\ 
			& & 0.6 & 0.66 & 2.42 & 1.02 & 0.95 \\ 
			& & 0.62 & 0.77 & 2.50 & 1.03 & 0.92 \\ 
			& & 0.63 & 0.69 & 2.11 & 0.98 & 1.09 \\ 
			\cline{2-7}
			 & \multirow{2}{*}{8} & 0.35 & 0.36 & 0.20 & 0.44 & 11.6 \\ 
			& & 0.385 & 0.5 & 0.85 & 0.72 & 2.70 \\ 
			\hline
		\end{tabular}
	\caption{For the injected capture waveforms with total mass $\Mtot$, mass ratio $q$ and initial angular momentum $L$ we write down the maximal redshift to which injections were made $z_{\max}$, the spacetime volume where the search is sensitive $\langle VT \rangle$, the sensitive distance reach $D_{\langle VT \rangle}$ and the rate upper limit $R_{90}$. The last three columns are for events detected with an iFAR $> 100$ years.}
	\label{tab:tabrange}
\end{table*}

\bibliographystyle{apsrev4-1}
\bibliography{references}

\end{document}